\documentclass[english]{article}
\usepackage{amsmath}
\usepackage{epsfig}
\usepackage[T1]{fontenc}
\usepackage[latin9]{inputenc}
\usepackage[letterpaper]{geometry}
\usepackage{amstext}
\usepackage[authoryear]{natbib}

\makeatletter
\newcommand{\lyxaddress}[1]{
\par {\raggedright #1
\vspace{1.4em}
\noindent\par}
}

\makeatother

\usepackage{babel}

\begin{document}

\title{Amino acid metabolism conflicts with protein diversity}

{\author{Teresa Krick$^{\text{}1}$, David A. Shub$^{\text{}2}$, Nina Verstraete$^{\text{}3}$,Diego U. Ferreiro$^{\text{}3}$,\\ 
Leonardo G. Alonso$^{\text{}4}$, Michael Shub$^{\text{}5}$ and Ignacio E. S\'anchez$^{\text{}3,*}$}}

\maketitle

\lyxaddress{1. Departamento de Matem\'atica, Facultad de Ciencias Exactas y
Naturales and IMAS - CONICET, Universidad de Buenos Aires.\\
2. Department of Biological Sciences, SUNY Albany.\\
3. Protein Physiology Laboratory, Departamento de Qu\'imica Biol\'ogica, Facultad de Ciencias Exactas y
Naturales and IQUIBICEN - CONICET, Universidad de Buenos Aires,
C1428EGA Buenos Aires.\\
4. Fundaci\'on Instituto Leloir - IIBBA CONICET, Buenos Aires. \\
5. IMAS - CONICET, Universidad de Buenos Aires.}

\lyxaddress{{*} Corresponding Author: Ignacio E. S\'anchez, isanchez@qb.fcen.uba.ar.}
\newpage{}

\section*{Abstract}

\paragraph*{The twenty protein coding amino acids are found in proteomes with different relative abundances. The most abundant amino acid, leucine, is nearly an order of magnitude more prevalent than the least abundant amino acid, cysteine. Amino acid metabolic costs differ similarly, constraining their incorporation into proteins. On the other hand, sequence diversity is necessary for protein folding, function and evolution. Here we present a simple model for a cost-diversity trade-off postulating that natural proteomes minimize amino acid metabolic flux while maximizing sequence entropy. The model explains the relative abundances of amino acids across a diverse set of proteomes. We found that the data is remarkably well explained when the cost function accounts for amino acid chemical decay. More than one hundred proteomes reach comparable solutions to the trade-off by different combinations of cost and diversity. Quantifying the interplay between proteome size and entropy shows that proteomes can get optimally large and diverse.}

\newpage{}

\section*{Introduction}

{T}he twenty proteinogenic amino acids are present in nature in different amounts, spanning nearly an order of magnitude (\citealp{uniprot}). The most abundant amino acid in both Swissprot and TrEMBL databases is leucine, while tryptophan and cysteine are the least abundant. According to statistical studies, natural protein sequences are indistinguishable from strings of amino acids chosen at random with the abovementioned abundances (\citealp{10988023}). Amino acid relative abundances are fairly well conserved across organisms, suggesting that a single underlying principle might determine the amino acid composition of proteomes.

Some forty years ago Dyer (\citealp{straw1, straw2}) suggested that protein sequences could be the result of transcription and translation of random DNA sequences. The  amino acid distribution arises from the interplay between the genomic GC content,  codon assignment and redundancy of the genetic code. We will refer to this as the genetic code model and describe it in more detail below. Despite its simplicity the calculated amino acid relative abundances correlate fairly well with the observed ones, although with prominent outliers (\citealp{straw1, straw2}).

The ``cost minimization principle" suggests that organisms minimize the cost of protein biosynthesis (\citealp{12574861,21604162}). A linear relationship between amino acid abundance and amino acid molecular weight or amino acid metabolic cost is supported by a reasonably high Pearson coefficient of correlation (\citealp{12574861,21604162}). However, the linear relationship is presented as such rather than justified from first principles (\citealp{12574861,21604162}) and cost minimization alone predicts that proteins would be homopolymers of the cheapest amino acid. On the other hand, natural protein folds can not be encoded with homopolymers, as described by the energy landscape theory of protein folding (\citealp{3478708}). A sufficiently large alphabet is needed to encode the diversity of known proteins (\citealp{9360595}). Precisely how  cost minimization and sequence diversity requirements balance each other is not known.

Here, we explicitly treat the trade-off between two competing forces: the minimization of the metabolic cost of amino acid biosynthesis and the maximization of the number of sequences that can be generated from a given amino acid composition. From this basic hypothesis, we deduce a mathematical relationship between amino acid metabolic cost and the logarithm of amino acid abundances. This simple relationship describes the data remarkably better than both the genetic code model and the linear cost-abundance model.

\section*{Theory}

\subsection*{A linear relationship}

A naive idea suggests that the probability that an amino acid is
incorporated in proteins might reflect the energetic cost of producing the
amino acid (with less costly amino acids used more frequently) while
maintaining the flexibility to code as many polypeptide chains as
possible. Previous work suggested that the relative abundance of
amino acids in proteomes is linearly related to the energetic costs
of making the amino acids (\citealp{12574861,21604162}). Here we suggest
that it is more appropriate to look for a linear relationship
between the logarithms of the relative abundances and the energetic costs.
We derive this relationship via a maximization principle.

Given probabilities $p_i,\ 1\le i\le 20$, representing the relative
abundances of the twenty amino acids, the number of probable peptide
chains of length $n$ can be calculated from Shannon's information theory as $e^{nh}$, where $$h=h(p_1,\dots,p_{20}) = -\sum_{i=1}^{20} p_i
\ln(p_i)$$ is the entropy (\citealp{Shannon1,Shannon}). The average energetic cost of amino acids in a cell is
 $\sum_{i=1}^{20} p_i e_i$, where $e_i$ is the energetic cost of
$i$-th amino acid.

The maximization of the number of sequences and the simultaneous minimization of metabolic cost is equivalent to maximizing the function
\begin{equation}\label{funcion}f(p_1,\dots,p_{20})=h(p_1,\dots,p_{20})-\sum_{i=1}^{20} p_i
e_i.\end{equation} The maximum of this function has the property that at a
given energetic cost the entropy is highest, that is the flexibility
to produce poly-peptide chains is greatest. Conversely, at a given entropy
the energy consumed by producing proteins is minimized. These properties hold for any choice of units for the energies and the entropy.

Maximizing $f$ predicts a linear relationship with negative slope
between the logarithms of the relative abundances and the energetic
costs. We maximize the function $f$ by differential calculus given a constraint, namely that the sum of the relative abundances equals unity, $\sum_{i=1}^{20} p_i
=1$. The gradient of the function should be a constant multiple of
the gradient of the constraint, the Lagrange
multiplier $\lambda$. Taking the partial derivative
with respect to $p_i$ of \eqref{funcion} and the constraint
$\sum_{i=1}^{20} p_i =1$ gives for each $i$:
$$-\ln(p_i) -1 - e_i = \lambda,\quad \mbox{ i.e. }\quad \ln(p_i)= -e_i - (1 +
\lambda).$$
The value of the intercept $-(1+\lambda)$ can
be derived from the constraint:
$$1=\sum_{j=1}^{20} p_j =\sum_{j=1}^{20}e^{-e_j - (1 +\lambda) } = e^{ - (1 + \lambda)}\sum_{j=1}^{20}e^{-e_j}$$
which implies that $-(1+\lambda)= -\ln(\sum_{j=1}^{20} e^{-e_j})$. This
gives the linear relation
\begin{equation}\label{relation}\ln(p_i)=
-e_i - \ln(\sum_{j=1}^{20} e^{-e_j}), \quad 1\le i\le
20,\end{equation}
between the logarithm of the relative abundance
and the energetic cost referred to above, with slope $-1$ when the
energetic cost $e_i$ is given in the ``correct" natural unit  e. Taking
the exponential of \eqref{relation}, this also yields the relative
abundance of the ith-amino acid $p_i$ in terms of the costs in unit
e:
\begin{equation}\label{freq}p_i= \frac{e^{-e_i}}{\sum_{j=1}^{20}
e^{-e_j}}.\end{equation} The formula is reminiscent of the Gibbs
distribution in physics.

\subsection*{The slope of the linear relationship}\label{slope}

Since the ``correct'' natural unit e for the energetic
costs $e_i$, $1\le i\le 20$ is not known, we can assume that the energetic
costs $c_i$ used in the examples below are given in terms of some
other unit c satisfying c$=m$e for some real $m$, and are
thus linear multiples of these theoretical $e_i$: $c_i=(1/m)e_i$ (or
$e_i=mc_i$) for $1\le i\le 20$. An important fact is that --under
the linear relationship derived in the previous section-- not only
is the relationship linear for this other choice of unit c (i.e.
for any other computed energetic cost), with slope $-m$ instead of
$-1$, but also the relative abundances $p_i$ are invariant under this
change of scale:

$$\ln(p_i)= -e_i - \ln(\sum_{j=1}^{20} e^{-e_j})= -m c_i - \ln(\sum_{j=1}^{20} e^{-mc_j}),$$ or equivalently
$$p_i=
\frac{e^{-e_i}}{\sum_{j=1}^{20}
e^{-e_j}}=\frac{e^{-mc_i}}{\sum_{j=1}^{20} e^{-mc_j}}, \quad 1\le
i\le 20.$$

In particular, if we use energetic costs $c_i$ measured in unit
c, and the observed slope in terms of this unit c is $-m$, then
letting e$=(1/m)$c we recover what we have called the ``correct"
natural unit e. We note that $1/m$ is analogous to the thermodynamic temperature in statistical mechanics. When we only have observed data, the slope of
the best fitting straight-line approximating the data may depend
on the scaling in some other way. That is if we multiply $e_i$ by
$1/m$ to get $c_i$, $1\le i\le 20$, the slope of the best linear
approximation may not multiply by $-m$. If it does multiply by $-m$
for all $m$ we say that the best straight-line approximation is
scale invariant. In this article, we use the reduced major axis (RMA) regression, which is scale invariant (Section Materials and Methods below). As such, the predicted relative abundances are independent of the scaling of the costs.

\section*{Results}\label{results}

\subsection*{Amino acid relative abundances in proteomes}\label{abundance}

We estimate amino acid relative abundances in
proteomes in two datasets. Dataset DS1 was derived from 108 fully
sequenced and annotated genomes from the three domains of life
(\citealp{17147802}). We translated coding regions into protein sequences
and counted the frequency of occurrence of each amino acid, assuming that all proteins are equally abundant (Table E1). Dataset DS2 was derived from the PaxDB database for protein abundances (\citealp{22535208}). We considered 17 organisms for which protein sequence and relative abundance data are available for more than 50 per cent of the proteome. We used integrated datasets for the whole
organism whenever possible (Table E2).

For both datasets, we tested several models for amino acid relative abundances. The results are shown in Table I, Figure 1 and Figure E1 below and explained
in the next sections.

\subsection*{Correlation of amino acid relative abundances with metabolic cost}\label{cost}

We test two linear relationships between amino acid
relative abundances and the metabolic cost, measured in ATP
molecules per molecule of amino acid. The first linear relationship correlates
(plain) relative abundances with costs, while the second one
correlates the logarithms of the relative abundances with costs.

We used the cost
estimation from (\citealp{11904428}), shown in Table II. Amino acid biosynthesis pathways are highly conserved across
organisms, as indicated by the high correlation between published
estimations of metabolic cost (\citealp{20808905}). 
Some organisms in DS1 and DS2 lack the biosynthetic
pathways for some amino acids, rendering them essential. If an amino
acid is essential, it is obtained from the environment and may be
then used for protein synthesis or catabolized. Similarly, if
an amino acid is not essential, it may or may not be produced by the
cell. The amount of energy that can be obtained from catabolizing an essential
amino acid is similar to the amount of energy that is needed for
its synthesis (\citealp{17476453}). Thus, the incorporation of
essential and non-essential amino acids in proteins involves similar energy
choices.

The plain amino acid relative abundances show a statistically
significant correlation with the amino acid metabolic cost (in ATP units) for
both datasets, with Pearson coefficients of correlation $r$ of -0.46
and -0.58 (Table I and Figure E1, panels A and C). The correlation is
also observed for individual organisms in DS1 and DS2 regardless of
genomic GC content (Figure 2, black lines in panels A and B). These
results are in agreement with previous proposals (\citealp{12574861,21604162}).

However, the theoretical model we put forward suggests that the correlation
should improve if we consider the logarithm of the amino acid relative
abundances instead of the relative abundances themselves. This is
indeed the case, as the $r$ values increase to -0.52 and -0.62 for
 DS1 and DS2 (Table I and Figure 1, panels A and D). The
correlation is better for most individual organisms in DS1 and DS2
regardless of genomic GC content (Figure 2, blue lines in panels A
and B). We conclude that our theoretical model describes the data
better than the previously reported empirical relationship between
amino acid costs and relative abundances.

\subsection*{Correlation of amino acid relative abundances with metabolic cost corrected by amino acid decay}\label{correlationdecay}

Amino acids undergo spontaneous chemical reactions in physiological
conditions and thereby degrade over time. Therefore, the metabolic
burden of amino acids should consider amino acid decay rates as well
as production cost. Since the experimental determination of the particular amino acid degradation rate is an extremely difficult task and we could not find a suitable set of amino acid decay rates in the literature, we have deduced a semi-quantitative reactivity ranking from previous publications and common knowledge of amino acid chemistry (described in detail in the Expandable text). We have taken into account nucleophilicity, redox reactivity and other biologically relevant reactions (\citealp{creighton}) (Table II). The physiological relevance of our ranking is supported by the presence of energy-consuming enzymatic pathways
that protect proteins against chemical decay (\citealp{17090414,22928493,9275166,12943218}).

Amino acid production cost and decay rates can be multiplied to
yield the amino acid production cost in units of ATP/time (Table II).
Plain amino acid production cost can be understood as the energy the
cell spends in making a molecule of a given amino acid. On the other
hand, this new quantity has units of power and can be understood as the
energy the cell spends per unit of time in order to keep a constant
concentration of a given amino acid, i.e., the energy flux through the metabolism of that amino acid (\citealp{16576642}).

We reassess the relationship between amino acid relative abundance and metabolic cost, as measured by energy flux in units of ATP/time. We observe a clearly improved correlation between
amino acid energy costs in units of ATP/time and both amino acid relative
abundances and their logarithms (Table I). In the case of the correlation with amino acid relative abundances, the $r$ values
increase to -0.72 and -0.79 for DS1 and DS2 (Figure E1,
panels B and D), regardless of genomic GC content (Figure
2, red lines in panels A and B). For the correlation with the
logarithm of amino acid relative abundances, the $r$ values further
rise to -0.86 and -0.91 for DS1 and DS2 (Figure 1,
panels B and E). The correlation is better for most
individual organisms in both datasets regardless of genomic GC
content (Figure 2, green lines in panels A and B). Thus, taking into account the simultaneous maximization of entropy and minimization of cost improves the correlation also when amino acid costs are measured in units of ATP/time.

The amino acid cysteine is very reactive, has a low relative abundance (empty symbols in Figures 1 and 4), a low cost in ATP units and a high cost in ATP/time units (Table II). Consequently, its relative abundance is
much better predicted when cost is considered in units of ATP/time
(Table I, Figure 1 and Figure 2). We have recalculated the
correlations for all models excluding cysteine in order to determine
whether the improvement in the $r$ values is due only to this
singular, very reactive amino acid (Table I). The main conclusions
of this work are valid for the remaining 19 amino acids as well. As
before, the $r$ value improves when we consider the logarithm of the
relative abundances instead of the relative abundances. Also, the
$r$ value increases when we consider amino acid costs in units of ATP/time.

We interpret that the proposed theoretical model, together with the
amino acid costs in units of ATP/time, is a very good descriptor
of amino acid relative abundances in proteomes. Compared with the
initial proposal of a linear relationship between amino acid
relative abundances and amino acid costs in units of ATP, the $r$ value improved
from -0.46 to -0.86 (DS1) and from -0.58 to -0.91 (DS2).

\subsection*{Correlation of amino acid relative abundances with the genetic code model}\label{codons}

The genetic code model relates amino acid relative abundance with the transcription and translation of random DNA sequences of a given GC content (\citealp{straw1, straw2}). To evaluate this model with DS1 and DS2 we retrieved the genomic GC content for each genome from (\citealp{22417913}) and used it to calculate the expected relative abundances for all 61 amino acid coding triplets. We then translated the triplets into amino acids and obtained the expected
amino acid relative abundances in each proteome. This metabolism-agnostic model shows a good correlation between calculated and observed amino acid relative abundances (Table I and Figure 1, panels C and F). The $r$ values are 0.71 and 0.62 for DS1 and DS2. The
correlation is also observed for individual organisms in the
database regardless of genomic GC content (Figure 2, dashed lines in
panels A and B). However, the $r$ values are worse than for our
metabolism-based model when amino acid costs are measured in units of ATP/time (Table I). This holds regardless of genomic GC content (Figure 2). The $r$ value is better for our model in 105 of the 108 organisms in DS1 (Figure 2, Panel A) and for the 17 organisms in DS2
(Figure 2, Panel B). This conclusion is also valid if the amino acid
cysteine is excluded from the calculations (Table I). We interpret
that amino acid relative abundances are better explained when we take into account the simultaneous maximization of entropy and minimization of cost.

\subsection*{The trade-off between amino acid metabolic cost and protein sequence diversity in natural proteomes}\label{tradeoff}

We postulate a model in which living organisms maximize a target function $f$ that equals the entropy of the amino acid distribution $h$ minus the average metabolic cost of an amino acid $\sum_{i=1}^{20} p_ie_im$. This gives rise to a trade-off between both terms. Figure 3 displays this trade-off for all organisms in DS1 (white symbols) and DS2 (black symbols). The figure also shows the expectation for the genetic code model (red symbols) and the expectation for the trade-off model (triangles). The figure plots the entropy $h$ of the amino acid distribution against the average amino acid metabolic cost in units of ATP/time. The contour lines indicate constant values of the target function $f$.

At a constant value of the entropy, most natural proteomes present lower metabolic costs than the genetic code model. Similarly, at a constant value of metabolic cost, most natural proteomes present higher entropies than the genetic code model. The target function $f$ takes higher values in most natural proteomes than in the genetic code model. Interestingly, each organism reaches the value of $f$ by a different combination of entropy and cost, with the costs varying as much as 20 per cent. The values of both entropy and cost lie within a restricted range. We interpret that the amino acid relative abundances in natural proteomes significantly deviate from the prediction of the genetic code model in a direction that simultaneously minimizes cost and maximizes sequence diversity, i.e, towards a better solution to the trade-off between metabolic cost and sequence diversity.

Most proteomes in DS1 have similar values of the target function $f$, while proteomes in DS2 show near-constant values of $f$. The values of $f$ are close to the expected values for the trade-off model calculated using equations 1 and 3, the costs in Table II and the values of $m$ for DS1 and DS2 from Figure 1B and 1E (triangles). This observation suggests that all organisms are close to a maximum in $f$, which is consistent with the maximization principle we have employed. At a maximum of $f$ the derivative is zero so the nearby values of the target function are nearly constant.

\section*{Discussion}\label{discussion}

Previous models for amino acid relative abundances in proteomes were based on the minimization of protein synthesis metabolic cost (\citealp{12574861,21604162}). However, the encoding and exploration of protein structure and function requires sequence diversity. We propose that the maximization of protein sequence diversity conflicts with cost minimization, determining proteome composition. The mathematical formulation of this concept gives rise to a trade-off that unites the two phenomena without introducing further priors and describes proteome composition with remarkable accuracy (Table I, Figure 1 and Figure 2).

Amino acids undergo spontaneous chemical reactions, as such the estimation of cost must take amino acid decay into account (Table II). We show that this leads to a more accurate description of amino acid distributions in proteomes (Table I). Consideration of both sequence diversity and amino acid turnover may also help in studying the relationship of amino acid metabolic cost with protein abundance (\citealp{11904428,17476453,22538926,18937004}), with amino acid substitution rates (\citealp{21604162,20808905}) and with the sequence properties of specific protein classes (\citealp{17684697,15853887,20824102,16762057}).

The model we put forward allows for a direct comparison between proteomes on a common basis (Figure 3). All natural proteomes fall along a line in the entropy-cost plane. This result arises from the observed amino acid relative abundances and the estimated metabolic costs and is independent of the mathematical shape of the relationship between abundances and costs. If the metabolic costs are organism-independent, this would indicate that there are multiple biological solutions to the entropy-cost trade-off. Some organisms have a lower average per amino acid cost and lower sequence diversity; while attaining higher sequence diversity is accompanied by a higher average per amino acid cost (Figure 3).

If the distribution of amino acids is equiprobable, the average metabolic cost per amino acid is 221 in units of ATP/time (Table II). For the average relative amino acid abundances in our datasets, the average metabolic cost drops to 129 in units of ATP/time. In other words, the metabolic cost of making a protein of length 100 for equiprobable amino acids is the same as the metabolic cost of making a protein of length 170 for the amino acid abundances in our datasets. Regarding sequence diversity, the number of probable proteins of length 100 is $e^{nh}$, where $h$ is the entropy. In the case of equally probable amino acids, $h\approx3.00$ nats and the number of probable proteins of length 100 is $\approx 10^{130}$. For the average relative amino acid abundances in our datasets, $h\approx 2.88$ nats. Comparing proteomes with the same average metabolic cost per amino acid, the number of probable proteins of length 170 is now $\approx 10^{212}$, with a gain of $\approx 10^{82}$ over the equiprobable case. Comparing proteins of length 100, the number of probable proteins for the average relative amino acid abundances in our datasets is $\approx 10^{125}$. The reduction on the possible proteins of length 100 by a factor of $10^5$ in natural proteomes relative to the equiprobable case might seem a sharp restriction. However, the $10^{125}$ remaining possibilities is far larger than the $10^{20}$ to $10^{50}$ sequences explored by terrestrial life since its origin (\citealp{18426772}). To sum up, we suggest that the cost-diversity trade-off allows for the efficient synthesis of large proteomes while not severely restricting protein diversification.

\section*{Materials and methods}

According to (\citealp[Table I5.1]{SoRo95}) and many other authors, we
chose to use here the reduced major axis (RMA) regression (or least
products regression) to fit the data, which is symmetric in both
variables, reflects better the best line fitting the data when both
variables are subject to errors and is scale invariant as mentioned
in Theory. The RMA regression computes the line $y =
m x + b$ for $m,b$ minimizing the function

$$f(m,b)=\sum_{i=1}^n \Big(y_i-(mx_i+b)\Big)\Big(x_i - (\frac{y_i-b}{m})\Big).$$

Denoting $\bar x=\frac{1}{n}\sum x_i$, $\bar y=\frac{1}{n}\sum y_i$
for the means, it is known that in our case

$$m=-\Big(\frac{\sum y_i^2 -n\bar y^2}{\sum x_i^2 - n\bar x^2}\Big)^{1/2} \quad \mbox{and} \quad
b=\bar y- m\bar x.$$

As usual, the Pearson product-moment correlation coefficient $r$,
$-1\le r\le 1$, given by the formula

$$r=\frac{\sum(x_i-\bar x)(y_i-\bar y)}{\sqrt{\sum_{i=1}^n(x_i-\bar x)^2}\sqrt{\sum_{i=1}^n(y_i-\bar y)^2}}
$$
(and satisfying that $r^2$ equals the usual $R^2$ coefficient of
determination), is used to measure how well the data fits the line:
in our case of negative slope, the closer $r$ is to $-1$ the better
it is.

\section*{Acknowledgements}

Teresa Krick and Michael Shub were supported by CONICET PIP 0801 2010-2012 and ANPCyT PICT 2010-00681. Ignacio E. S\'anchez was supported by ANPCyT PICT 2010-1052. Nina Verstraete is the recipient of a CONICET postdoctoral fellowship. We would like to thank Shuai Cheng Li and Lu Zhang, from Hong Kong City University, for their help and Raik Gruenberg, Thierry Mora, Pedro Beltrao and Jesus Tejero for discussion.

\section*{Author Contributions}

MS conceived the project. MS, IES, DAS, TK and DUF developed the idea and designed the experiments. IES, TK, LGA and NV performed the experiments. MS, IES, TK and DUF analyzed the results. MS, TK, DUF and IES wrote the paper.

\section*{Additional Expanded View Figure Legends}

\paragraph*{Table~E1 }

Dataset DS1 was derived from 108 fully
sequenced and annotated genomes from the three domains of life
(\citealp{17147802}). Coding regions were translated into protein sequences
and the frequency of occurrence of each amino acid was calculated, assuming that all proteins are equally abundant. The table shows values of amino acid relative abundances, predicted abundances from the genetic code model, genomic GC content and correlation R-values.

\paragraph*{Table~E2 } 

Dataset DS2 was derived from 17 organisms from the PaxDB database for protein abundances (\citealp{22535208}). We considered organisms for which protein sequence and relative abundance data are available for more than 50 per cent of the proteome and used integrated datasets for the whole
organism whenever possible. The table shows values of amino acid relative abundances, predicted abundances from the genetic code model, genomic GC content and correlation R-values.

\paragraph*{Expandable text} Deduction of a semi-quantitative reactivity ranking for the twenty proteinogenic amino acids.

\newpage

\section*{Tables}

\paragraph*{Table~I }

Pearson's correlation coefficients for correlation of amino acid relative abundances with amino acid metabolic cost and a model based on the genetic code. 
The two columns labeled with (no C) are the results of the same calculations excluding the amino acid cysteine.

\begin{table*}[ht]
\noindent {\footnotesize{
\begin{tabular} [c]{|c|c|c|c|c|c|}
 \hline {\bf Model} & {\bf DS1} & {\bf DS1 (no C)} & {\bf DS2} & {\bf DS2 (no C)}
\\ \hline {\bf Cost(ATP) vs. abundance} & -0.46 & -0.51 & -0.58 & -0.64
\\ \hline {\bf Cost(ATP) vs. ln(abundance)} & -0.52 & -0.64 & -0.62 & -0.75
\\ \hline {\bf Cost(ATP/time) vs. abundance} & -0.72 & -0.68 & -0.80 & -0.76
\\ \hline {\bf Cost(ATP/time) vs. ln(abundance)} & {\bf -0.86} & {\bf -0.83} & {\bf -0.91} & {\bf -0.90}
\\ \hline {\bf Genetic code model vs. ln(abundance)} & 0.71 & 0.76 & 0.62 & 0.66
\\ \hline
\end{tabular}
}}
\end{table*}

\paragraph*{Table~II }

Amino acid metabolic cost. Costs in units of ATP molecules
per amino acid molecule are from (\citealp{11904428}), costs in units of
ATP molecules per amino acid molecule corrected by amino acid decay
are from this work. The estimation of amino acid reactivity and
decay rates (in relative units) is described in the expandable material.

\begin{table*}[ht]
\noindent {\small{
\begin{tabular} [c]{|c|c|c|c|c|}
\hline {\bf Amino} & Cost & Decay &
{\bf Cost }
\\[-1mm] {\bf acid } & (ATP) & (1/time) &{\bf(ATP/time)}
\\ \hline {\bf A} & 11.7 & 1 & {\bf 12 }
\\ \hline {\bf C} & 24.7 & 30 &{\bf 741 }
\\ \hline {\bf D} & 12.7 & 9 &{\bf 114 }
\\ \hline {\bf E} & 15.3 & 5 &{\bf 77 }
\\ \hline {\bf F} & 52 & 4 &{\bf 208 }
\\ \hline {\bf G} & 11.7 & 1 &{\bf 12 }
\\ \hline {\bf H} & 38.3 & 14 &{\bf 536 }
\\ \hline {\bf I} & 32.3 & 2 &{\bf 65 }
\\ \hline {\bf K} & 30.3 & 8 &{\bf 242 }
\\ \hline {\bf L} & 27.3 & 2 &{\bf 55 }
\\ \hline {\bf M} & 34.3 & 13 &{\bf 446 }
\\ \hline {\bf N} & 14.7 & 10 &{\bf 147 }
\\ \hline {\bf P} & 20.3 & 3 &{\bf 61 }
\\ \hline {\bf Q} & 16.3 & 8 &{\bf 130 }
\\ \hline {\bf R} & 27.3 & 4 &{\bf 109 }
\\ \hline {\bf S} & 11.7 & 6 &{\bf 70 }
\\ \hline {\bf T} & 18.7 & 6 &{\bf 112 }
\\ \hline {\bf V} & 23.3 & 2 & {\bf 47 }
\\ \hline {\bf W} & 74.3 & 12 &{\bf 892 }
\\ \hline {\bf Y} & 50 & 7 & {\bf 350 }
\\ \hline
\end{tabular}
}}
\end{table*}

\newpage

\begin{figure*}[hbtp]
\begin{center}
\centerline{\includegraphics[width=.80\textwidth]{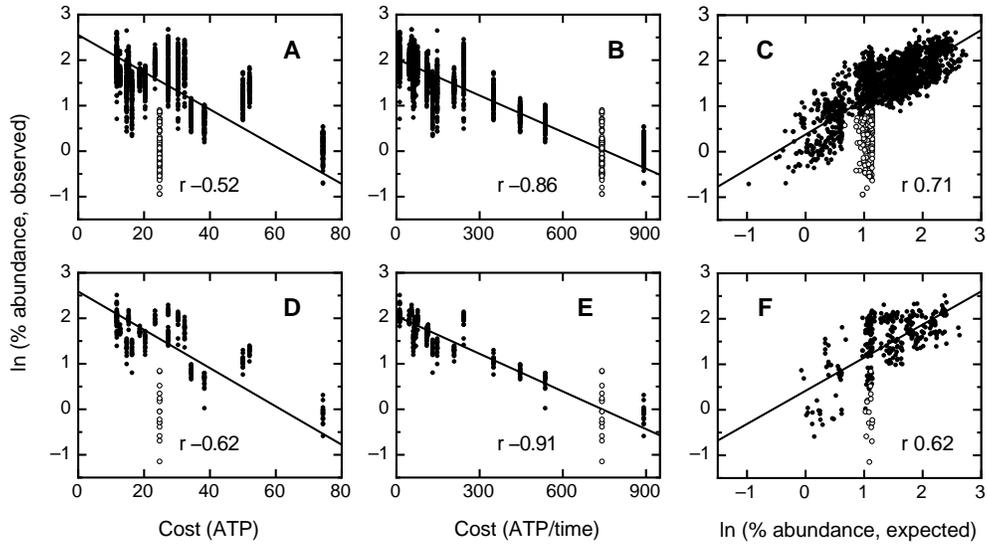}}
\caption{Amino acid metabolic cost corrected by amino acid decay explains amino acid relative abundances better than uncorrected amino acid metabolic cost and a model based on the genetic code. 
Panels A, B and C correspond to Dataset DS1. Panels D, E and F correspond to Dataset DS2. Amino acid metabolic costs are shown in units of ATP molecules per amino acid molecule (panels A and D) and in units of ATP molecules per amino acid molecule corrected by amino acid decay (panels B and E). Panels C and F represent correlation between observed and expected amino acid relative abundances (genetic code model). Data points for the amino acid cysteine are shown as empty symbols, the rest of the amino acids are shown as black symbols. The lines are RMA regressions to all data points.} \label{Figure1}
\end{center}
\end{figure*}

\newpage

\begin{figure*}[hbtp]
\begin{center}
\centerline{\includegraphics[width=.80\textwidth]{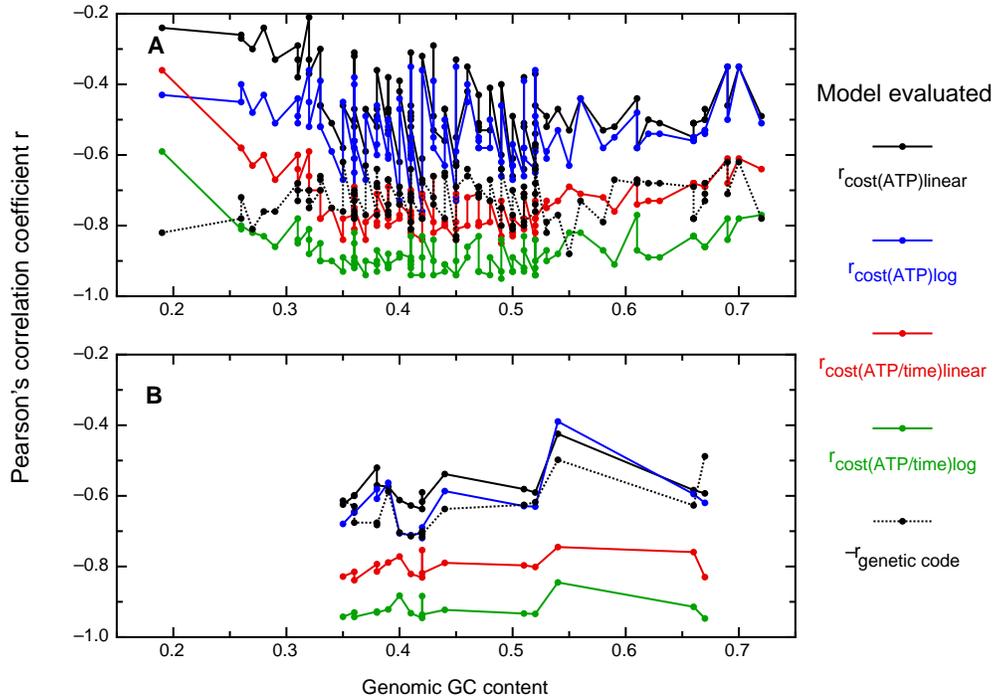}}
\caption{Correction for amino acid decay and use of logarithmic values improves the correlation between amino acid relative abundances and amino acid metabolic cost, independently of genomic GC content. 
Panel A corresponds to Dataset DS1, panels B corresponds to Dataset DS2. 
Pearson's correlation coefficients for correlation of amino acid relative abundances with amino acid metabolic costs are plotted for each proteome of the corresponding dataset. 
Amino acid costs were calculated as units of ATP molecules per amino acid (black and blue lines) or as units of ATP molecules per amino acid molecule corrected by amino acid decay (red and green lines). Amino acid relative abundances values were taken as plain or logarithmic values (black/red lines and blue/green lines, respectively). 
Dashed lines correspond to correlation coefficient values for correlation between observed and expected amino acid relative abundances (genetic code model). The data are shown as a function of genomic GC content in the x axis.} \label{Figure2}
\end{center}
\end{figure*}

\newpage

\begin{figure*}[hbtp]
\begin{center}
\centerline{\includegraphics[width=.80\textwidth]{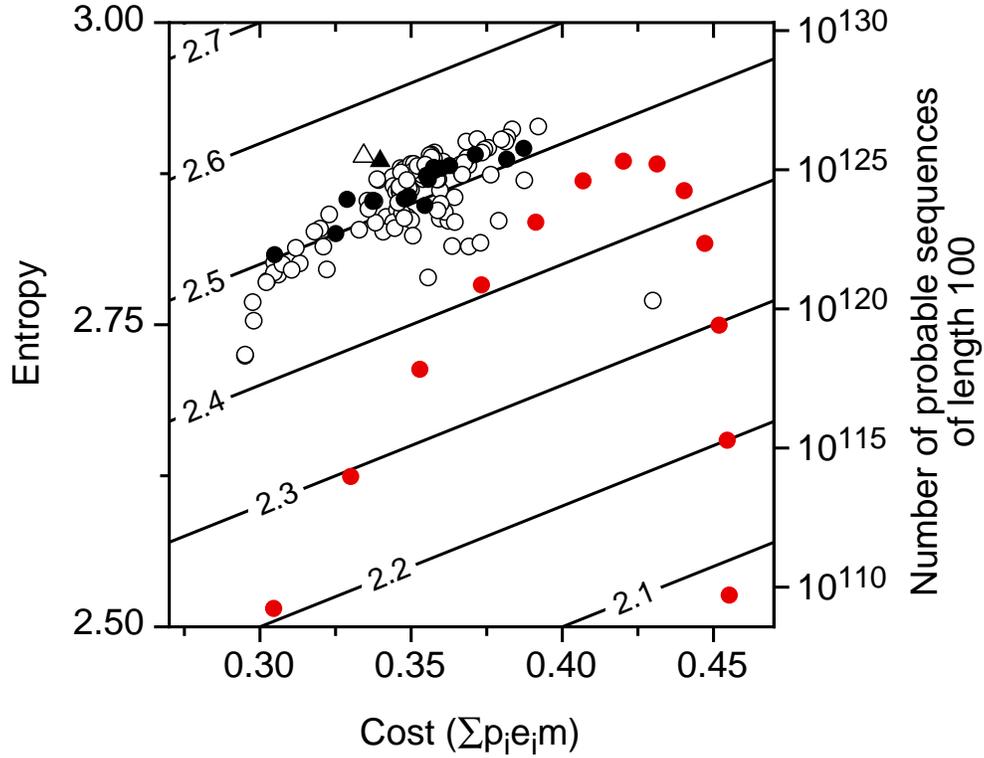}}
\caption{Natural proteomes reach comparable solutions to the trade-off between amino acid metabolic cost and sequence diversity. 
Amino acid metabolic cost (x-axis) and protein sequence diversity measured as entropy (y-axis) are plotted for the 107 organisms in Dataset DS1 (white symbols) and the 17 organisms in Dataset DS2 (black symbols). Values for the genetic code model (red symbols) are plotted for genomic GC contents between 0.15 (lower right corner) to 0.75 (lower left corner). Triangle symbols represent values for the trade-off model using the values of m (slope of the linear relationship) for DS1 and DS2 from Figure 1B and 1E, respectively. The contour lines indicate the value for the target function f. The y-axis legend to the right illustrates the number of probable peptide chains of length 100 given by $e^{100h}$, where $h$ is the entropy (\citealp{Shannon1, Shannon}).} \label{Figure3}
\end{center}
\end{figure*}

\newpage

\begin{figure*}[hbtp]
\begin{center}
\centerline{\includegraphics[width=.80\textwidth]{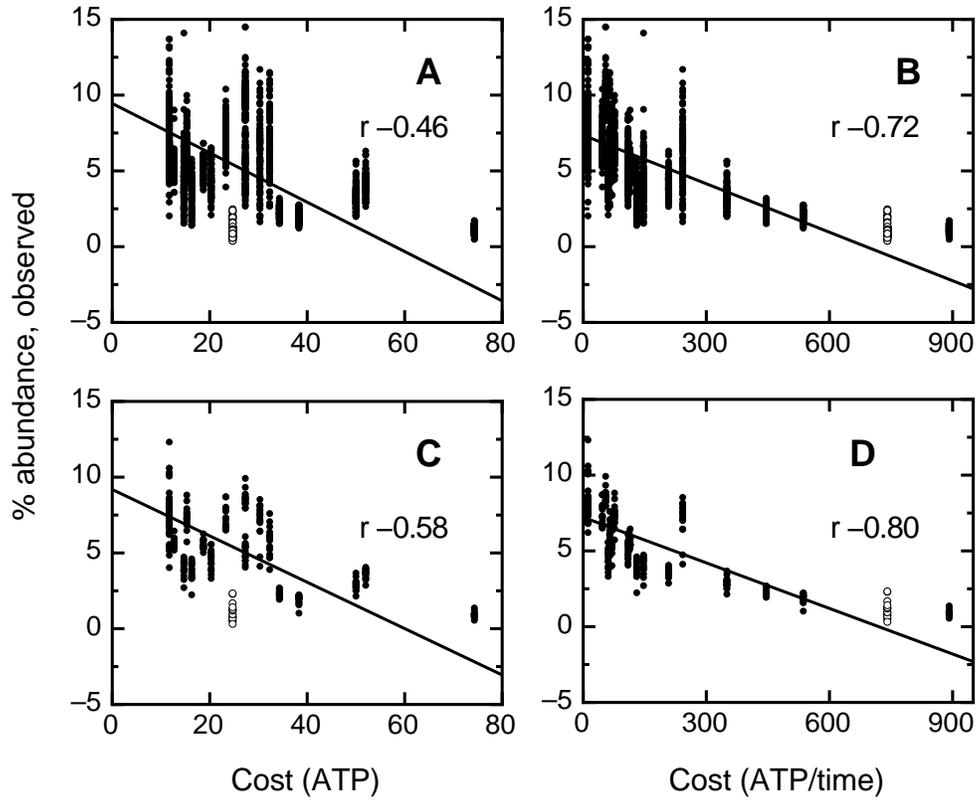}}
\caption{Amino acid metabolic cost corrected by amino acid decay explains amino acid relative abundances better than uncorrected amino acid metabolic cost. 
Panels A and B correspond to Dataset DS1. Panels C and D correspond to Dataset DS2. Amino acid metabolic costs are shown in units of ATP molecules per amino acid molecule (panels A and C) and in units of ATP molecules per amino acid molecule corrected
by amino acid decay (panels B and D). Data points for the amino acid cysteine are shown as empty symbols, the rest of the amino acids are shown as black symbols. The lines are RMA regressions to all data points.} \label{FigureE1}
\end{center}
\end{figure*}

\newpage

\bibliographystyle{msb}
\bibliography{NaiveIdea-MSB-7-03-14}

\end{document}